\documentclass{article}

\AtBeginDocument{%
  }

\usepackage{authblk}
\usepackage{hyperref}
\usepackage{paralist}
\usepackage{placeins}
\usepackage{amsmath,amsfonts,amsthm}
\usepackage{multirow,multicol}
\usepackage{booktabs}
\usepackage{pdflscape}
\usepackage{graphicx}
\usepackage{caption}
\usepackage{subcaption}
\usepackage{xspace}
\usepackage{enumerate}

\usepackage{tikz}
\usepackage{pgfplots}
\usepackage{pgfmath}
\usetikzlibrary{plotmarks,shapes,pgfplots.dateplot}
\usetikzlibrary{math,calc,trees,positioning,arrows,chains,shapes.geometric,%
    decorations.pathreplacing,decorations.pathmorphing,shapes,%
    matrix,shapes.symbols}
\usetikzlibrary{3d,decorations.text,shapes.arrows,fit,backgrounds,quotes,arrows.meta}
\usepackage{ifthen}
\usepackage{numprint}

\begin{document}

\title{STIR: Siamese Transformer for Image Retrieval Postprocessing}
\date{\today}

\author[1]{Aleksei Shabanov}
\author[2]{Aleksei Tarasov}
\author[1]{Sergey Nikolenko}
\affil[ ]{
\{shabanoff.aleksei,  aleksei.v.tarasov\}@gmail.com,  sergey@logic.pdmi.ras.ru
}
\affil[1]{St. Petersburg Department of the Steklov Institute of Mathematics, St. Petersburg, Russia}
\affil[2]{New Yorker  GmbH, Berlin, Germany}


\maketitle

\begin{abstract}
Current metric learning approaches for image retrieval are usually based on learning a space of informative latent representations where simple approaches such as the cosine distance will work well. Recent state of the art methods such as HypViT move to more complex embedding spaces that may yield better results but are harder to scale to production environments. In this work, we first construct a simpler model based on triplet loss with hard negatives mining that performs at the state of the art level but does not have these drawbacks. Second, we introduce a novel approach for image retrieval postprocessing called Siamese Transformer for Image Retrieval (STIR) that reranks several top outputs in a single forward pass. Unlike previously proposed Reranking Transformers, STIR does not rely on global/local feature extraction and directly compares a query image and a retrieved candidate on pixel level with the usage of attention mechanism. The resulting approach defines a new state of the art on standard image retrieval datasets: Stanford Online Products and DeepFashion In-shop. We also release the source code\footnote{
https://github.com/OML-Team/open-metric-learning/tree/main/pipelines/postprocessing/
} and an interactive demo\footnote{https://dapladoc-oml-postprocessing-demo-srcappmain-pfh2g0.streamlit.app/} of our approach.
\end{abstract}

\section{Introduction}

Modern approaches for metric learning and image retrieval usually employ a standard pretrained backbone which is fine-tuned for the task with a metric learning objective such as the triplet loss~\cite{10.5555/1756006.1756042}. Much of the progress in the field has concentrated on ways to improve upon the basic triplet loss. The backbones have usually been standard successful deep architectures, first convolutional ones such as ResNet-50 and later Transformer-based such as the Vision Transformer (ViT)~\cite{DBLP:journals/corr/abs-2010-11929}.

\begin{figure}[!t]
    \centering
    \includegraphics[width=\linewidth]{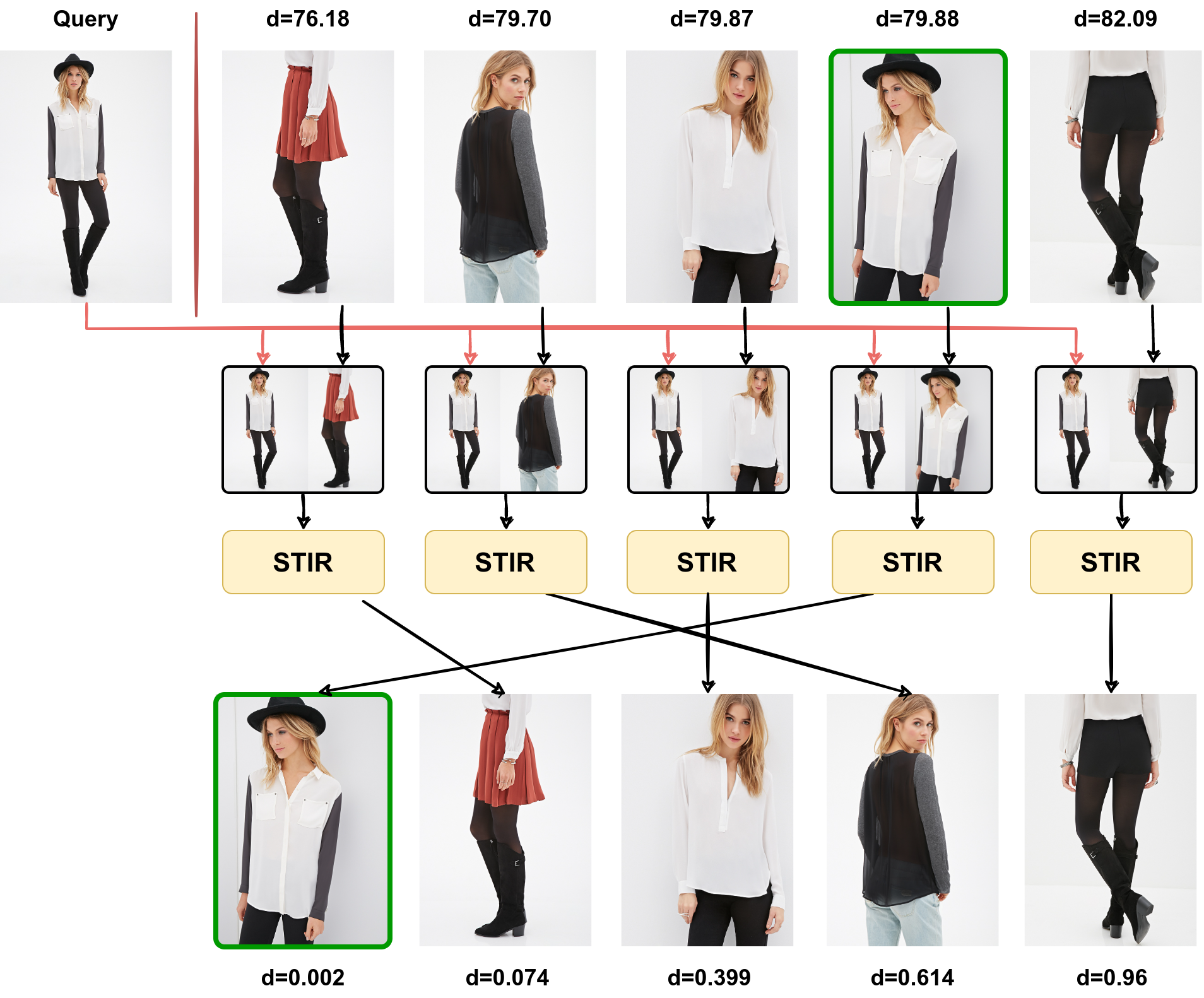}
    \caption{STIR postprocessing on a real example from the In-Shop dataset: the reranking model receives as input concatenated query and gallery images and outputs the probability of them being a negative pair.}\label{fig:sample}
\end{figure}

In 2020, Musgrave at el.~\cite{10.1007/978-3-030-58595-2_41} performed an experimental evaluation of a long line of metric learning results and found little improvement over standard approaches, concluding that (undeniable) progress had been mostly due to steadily improving backbones. Since then, metric learning and information retrieval have become dominated by Transformer-based architectures, with ViT~\cite{DBLP:journals/corr/abs-2010-11929} being especially influential for image retrieval. The standard baseline today is a ViT-like model fine-tuned with the triplet loss to produce a latent space where the dot product of embeddings corresponds to the similarity needed for the retrieval problem. Still, latest works claim significant improvements with a variety of new loss functions~\cite{NEURIPS2021_c622c085,patel2022recall} and even remapping the embeddings into a hyperbolic space~\cite{9880306} (see Section~\ref{sec:related}).

Our first contribution in this work is to go back and re-evaluate the standard triplet loss-based approach with a ViT backbone, which we call \emph{ViT-Triplet} (Fig.~\ref{fig:stir}a). We find that with a brief tuning of hyperparameters and efficient implementation, ViT-Triplet outperforms state of the art results in some settings and reaches similar results in others. Apart from improved results, ViT-Triplet, in our opinion, is a better option in practice since other solutions are either harder to bring to production environments or require more complicated tuning.

Second, we consider postprocessing for ViT-Triplet in the form of reranking the top results. Reranking for image retrieval has a long history~\cite{DBLP:journals/corr/NohASH16,8954470,10.1007/978-3-030-58565-5_43,sarlin20superglue,DBLP:conf/iccv/TanYO21} but it has usually been applied to relatively weak models, where one needed to rerank hundreds of results. We consider reranking for ViT-Triplet output and note that since the results of ViT-Triplet are already quite good we can concentrate on reranking the top few results, which allows us to use much more computationally intensive methods. We present the \emph{Siamese Transformer for Image Retrieval} (STIR) model that uses a ViT architecture to process a concatenation of each query-result pair with a small MLP head on top (Fig.~\ref{fig:stir}b). We show that STIR indeed improves over ViT-Triplet and prior art and thus sets new state of the art for several well-known image retrieval datasets. Fig.~\ref{fig:sample} illustrates sample STIR reranking on the In-Shop dataset~\cite{liuLQWTcvpr16DeepFashion}: distances on top are the result of ViT-Triplet, distances at the bottom are the results of STIR, and the ground truth answer is highlighted in green.

The paper is organized as follows: Section~\ref{sec:related} reviews related work, Section~\ref{sec:method} introduces ViT-Triplet and STIR reranking, Section~\ref{sec:eval} presents our evaluation results, and Section~\ref{sec:concl} concludes the paper.

\section{Related work}\label{sec:related}

We identify two relevant directions of related work. 
First, image retrieval itself, where the best recent work employs Transformer-based backbones. Vision Transformers~\cite{DBLP:journals/corr/abs-2010-11929} were fine-tuned for image retrieval by the IRT model~\cite{DBLP:journals/corr/abs-2102-05644} based on the DeiT distillation approach~\cite{pmlr-v139-touvron21a}. Hyperbolic Vision Transformers (Hyp-ViT)~\cite{9880306} reach state of the art results on several datasets by using pairwise cross-entropy with hyperbolic distances measured on the Poincar{\'e} ball. However, the resulting embeddings are not suitable for most existing vector search engines that rely on algorithms optimized for Euclidean spaces, so Hyp-ViT is hard to bring to production environments. Moreover, Hyp-ViT defines an entire family of models (six variations, two embedding sizes for each) that need to be evaluated in each case, which we view as a kind of hyperparameter tuning.

Another direction of study introduces new loss functions that approximate or provide bounds for non-differentiable retrieval metrics. ROADMAP~\cite{NEURIPS2021_c622c085} presents a decomposable differentiable upper bound for the average precision. Patel et al.~\cite{patel2022recall} proposed a differentiable surrogate loss for recall optimization further augmented with mixup regularization; this, however, also leads to a set of new hyperparameters such as sigmoid temperatures that need to be tuned. We also note the HAPPIER model that proposes a new loss function for hierarchical image retrieval~\cite{ramzi2022hierarchical}. Interestingly, despite the prevalence of Transformers some of the top results are still produced by CNNs: a combination of multiple CNN-based global descriptors was proposed in~\cite{DBLP:journals/corr/abs-1903-10663}, while the standard ResNet-50 backbone has been leveraged with the ProxyNCA++ method (an update on proxy-neighborhood component analysis) in~\cite{10.1007/978-3-030-58586-0_27} and with the Metrix loss function that extends mixup to metric learning objectives (including triplet loss) in~\cite{venkataramanan2022it}.

Second, specifically postprocessing (reranking) approaches are rare in recent works. Classical approaches usually reranked image retrieval results based on local descriptors extracted from the images~\cite{DBLP:journals/corr/NohASH16,8954470,10.1007/978-3-030-58565-5_43}. We note \emph{SuperGlue} that used graph neural networks to link local descriptors~\cite{sarlin20superglue} and the \emph{Reranking Transformer} (RRT) approach that uses a Transformer to process global and local descriptors extracted from an image pair~\cite{DBLP:conf/iccv/TanYO21}. Unlike most approaches that rerank a large set of results (at least several hundred), we aim to correct an already high-performing Transformer-based model so we concentrate on (relatively heavyweight) postprocessing of a few top results.

\section{Method}\label{sec:method}

\subsection{ViT-Triplet}

For the ViT-Triplet model, we follow the approach from~\cite{DBLP:journals/corr/HermansBL17,DBLP:journals/corr/abs-1912-07863} and fine-tune a ViT backbone for image retrieval as shown in Fig.~\ref{fig:stir}a. We form batches by taking $P$ labels (item ids) and $K$ instances (images) for each label. 
To decrease the number of hyperparameters we set $P=4$ since the median size of a class is 5 in InShop and 4 in SOP (see Section~\ref{sec:data}), so we can avoid severe under- or oversampling. The parameter $K$ is chosen such as to fill the GPU memory ($K=150$ in our case for an NVIDIA V100 GPU). 


After a batch is sampled, we perform hard triplet mining to form $PK$ triplets; namely, we calculate the distance matrix between embeddings of images in the batch and take for each image the hardest positive sample (same label, maximum distance) and the hardest negative sample (different label, minimum distance). 

Then we compute the triplet loss function 
$$L(q,p,n) = \max\left(0, d(q,p) - d(q,n) + m\right)$$
for a query image $q$, positive sample $p$, negative sample $n$, distance in the embedding space $d(\cdot,\cdot)$, and constant margin $m$; we set $m=0.15$ in all experiments (in previous works, it was usually chosen as $m\in [0.1, 0.2]$). We name the resulting model \emph{ViT-Triplet}.


\subsection{Siamese Transformer}

Qualitative error analysis has shown that many mistakes may be caused by the fact that the feature extractor has to ``blindly'' represent a given image as a vector in the latent space, without understanding what other images it would be compared against. In a perfect world, all the information needed for this comparison would be already included in the feature vector, but in reality, the model can benefit a lot from a direct side-by-side comparison of the images. 

Another motivation comes from the distribution of results. For instance, the CMC@1 metric for \emph{ViT-Triplet} on the In-Shop dataset is 92.1\% but CMC@5 reaches 97.6\%, i.e., less than a third of the error rate. It means that we usually already have a 
correct answer at the top of the list, and side-by-side comparison may help us push it in the first place. In general, with a good feature extractor CMC@k saturates for relatively small values of $k$, such as $k=5$, so we do not need to rerank a lot and can afford to use a relatively heavyweight model.

We suggest to use a reranking model that performs pairwise comparisons of the query image and top retrieved results. We want our pairwise postprocessor to have the following properties:
\begin{enumerate}[(i)]
    \item it has to reuse an already trained feature extractor, ideally without new large trainable networks;
    \item it has to have an attention mechanism to compare the regions of a query image and regions of a gallery image pairwise;
    \item it has to be interpretable or at least provide some mechanism that can be used to interpret the results;
    \item it has to be simple and not require additional manual labeling or extra data.
\end{enumerate}

\begin{figure}[!t]
    \centering
    \includegraphics[width=.7\linewidth]{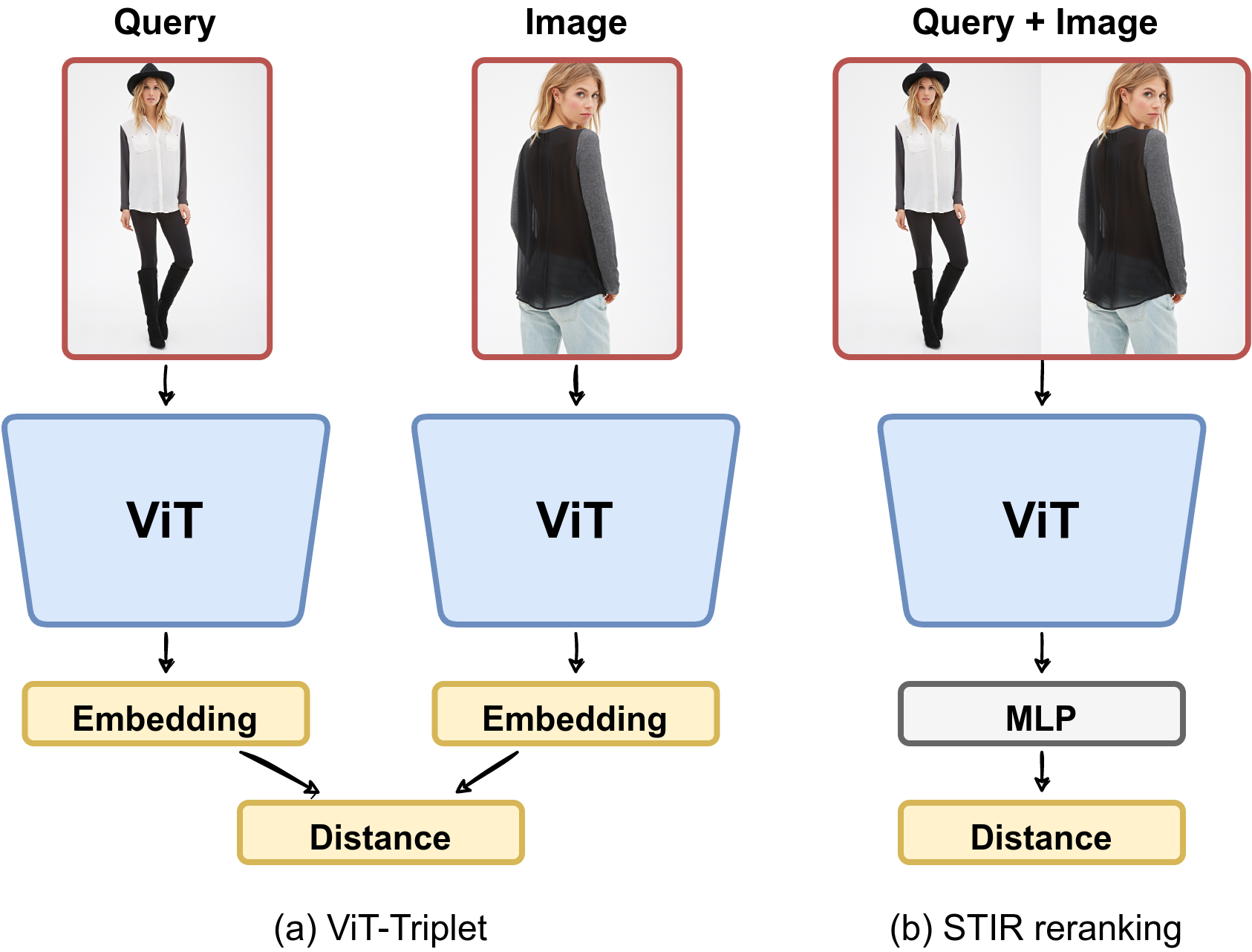}
    \caption{Network architectures: (a) ViT-Triplet; (b) STIR postprocessing.}\label{fig:stir}
\end{figure}

Thus, we propose to use the \emph{Siamese Transformer for Image Retrieval} (STIR) model, which is a ViT feature extractor with an additional MLP on top that takes two concatenated images as input and returns the probability of these images to be a negative pair (Fig.~\ref{fig:stir}b). This output can also be interpreted as a ``distance'': lower probability means more similar images. STIR satisfies the requirements above:
\begin{enumerate}[(i)]
    \item a pretrained \emph{ViT-Triplet} is used for initialization;
    \item the built-in ViT attention mechanism considers the interactions of patches both inside an image and across images;
    \item the resulting attention maps help to achieve interpretability;
    \item the only overhead is a two-layer MLP, and the input can reuse the same image pairs.
\end{enumerate}

To train the postprocessor, similarly to \emph{ViT-Triplet} we form batches by taking $P$ labels and $K$ images for each label, with the same $K=4$ but $P=30$ instead of $150$ since STIR has a larger memory footprint due to larger input size. After a batch is sampled, we mine hard pairs (pairs with largest distances and same labels or smallest distances and different labels), concatenate the images, and feed them to STIR, which predicts the
probability of a pair to be negative.
We use the binary cross-entropy as the objective function for STIR.

Another important property of STIR is that it is \emph{asymmetric}, i.e., the results may depend on whether we put the query on the left and a gallery image on the right or vice versa. We have not found significant differences between these two options, but the results improve a little further if we symmetrize STIR by averaging their results. We call this version \emph{STIR-Symmetric} in the tables below and propose it as a slightly improved version of STIR reranking with an additional cost of running the model twice.

\section{Evaluation}\label{sec:eval}

\subsection{Datasets and experimental setup}\label{sec:data}

We concentrate on two standard image retrieval datasets.

The \emph{In-shop Clothes Retrieval Benchmark} (\textbf{In-Shop}) dataset~\cite{liuLQWTcvpr16DeepFashion} is a part of the \emph{DeepFashion} dataset with \numprint{7982} clothing items and \numprint{52712} high quality in-shop images, with the median of $5$ photos per item. There are \numprint{25882} images in the training set and \numprint{26830} images in the test set, which is divided into two non-overlapping parts: query set (\numprint{14128} images) and gallery (the search index, \numprint{12612} images). Each query image corresponds to one or more images of the same clothing item in the gallery. 

\emph{Stanford Online Products} (\textbf{SOP})~\cite{song2016deep} has \numprint{22634} online products with \numprint{120053} related images, with the median of $4$ photos per item. There are \numprint{11318} products (\numprint{59551} images) in the training set and \numprint{11316} products (\numprint{60502} images) in the test set. The test set has no fixed query-gallery split, so we consider each individual photo as a query and evaluate it versus the rest of the images in the test set.

To ensure a fair comparison, we copy (as much as possible) the parameters and backbone models for ViT-Triplet from Hyp-ViT~\cite{9880306}. The model architecture is ViT-S/16 (small version, patch size 16), the optimizer is AdamW with lr=1e-5, the image size is 224, and augmentation transforms are Horizontal Flip and Random Resized Crop with scale randomly chosen from $(0.2, 1.0)$; we did not use any additional information from the data such as bounding boxes or category labels.
The training setup for STIR is mostly the same as for ViT-Triplet: image size 224, the same augmentations, but with a less agressive Random Resized Crop (sampling the scale parameter from $(0.8, 1)$) so that STIR can compare almost the entire two images side-by-side. We used the AdamW optimizer with learning rate 2e-3 for the first 3 epochs, when we fine-tune the MLP head only, and 1e-5 for the rest of the training. The MLP head consists of two fully connected layers with sizes (384, 192) and (192, 1) respectively, separated by a dropout layer with probability $p=0.5$ and sigmoid activation function.
We run all training experiments on two NVIDIA V100 GPUs with half-precision turned on. For the final metrics evaluation, we used only one GPU and turned off half-precision.

In the evaluation tables, all external results are taken from the corresponding papers except for surrogate recall, where the original work~\cite{patel2022recall} reports only the ViT-B version, which is better than the results in Table~\ref{tbl:res} that use the ViT-S backbone. Therefore, we have re-evaluated surrogate recall with ViT-S using the original code~\cite{patel2022recall}.

\subsection{Evaluation metrics}

\def\ntopk{n_{\mathrm{k}}}
\def\ngt{n_{\mathrm{gt}}}

Most works on metric learning and information retrieval report Recall@k for various values of $k$ as the primary evaluation metric. Interestingly, there is a significant discrepancy between the understanding of recall in classical information retrieval and the ``recall'' metric used in many works on metric learning. 

Usually, recall is defined as 
$$\mathrm{Recall}@k = \frac{\ntopk}{\ngt},$$ 
where $\ntopk$ is the number of ground truth results in the top $k$ retrieved results and $\ngt$ is the total number of ground truth results. However, metric learning works often report, e.g., Recall@1 values close to $1$ even when there exist several ground truth answers to a query, $\ngt > 1$, and Recall@1 should be bounded by $1 / \ngt$. This is because instead of recall they actually report the \emph{cumulative matching characteristics} (CMC) metric: 
$$\mathrm{CMC}@k=\begin{cases} 1, & \text{if a correct answer is among top $k$ retrieved results,} \\  0, & \text{otherwise}.\end{cases}$$

For datasets with exactly one ground truth answer for every query, CMC and recall coincide; however, this is not the case for In-Shop and SOP datasets so we keep the CMC terminology in evaluation tables. We also note that since In-Shop and SOP have several correct answers, the \emph{precision} metric, 
$$\mathrm{Precision}@k = \frac{\ntopk}{k},$$ also makes sense for evaluation. Therefore, below we report mean average precision (mAP) values as well, where
$$\mathrm{AP@k} = \frac{1}{\ntopk}\sum_{i=1}^k\left[\#i\text{ is correct}\right]\cdot\mathrm{Precision}@i$$ 
is the average precision (area under the precision-recall curve), and mAP@k is AP@k averaged over the test set queries.  
Unfortunately, we have nothing to compare with in terms of mAP since its values have not been reported in previous works.

\subsection{Results}

Table~\ref{tbl:res} shows the main results of our comparison. First, note that the ViT-Triplet model, trained with the standard embedding dimension $384$ only with the triplet loss and ViT backbone, shows state of the art results on both SOP and In-Shop datasets, losing to the current state of the art HypViT only in the CMC@1 metric on In-Shop. This supports our conclusion that most of the latest progress in image retrieval has been due to steadily improving Transformer-based backbones, and a well-trained ViT backbone with a straightforward triplet loss is still a very competitive approach to image retrieval.

Second, Table~\ref{tbl:res} shows how STIR postprocessing improves the results of ViT-Triplet, outperforming the best previous results (including ViT-Triplet itself) and Reranking Transformers in the CMC@1 metric (Recall@1). Note that since STIR in Table~\ref{tbl:res} is limited to reranking the top $n=5$ results, it cannot change the CMC@k and Recall@k 
metrics for $k\ge 5$, so the rest of the results coincide with ViT-Triplet. STIR results improve monotonically with $n$, but we have chosen $n=5$ to report in Table~\ref{tbl:res} because in this case, STIR postprocessing has the same running time as reranking Transformers~\cite{DBLP:conf/iccv/TanYO21}.

In Table~\ref{tbl:map}, we report mean average precision scores for our methods; we do not have the results of other approaches here, so we show these numbers to provide a baseline and hope that later works will measure mAP as well.

\begin{table*}[p]\centering
\caption{Image retrieval results on SOP and In-Shop datasets, CMC metric; best results highlighed in bold.}\label{tbl:res}
\setlength{\tabcolsep}{5pt}
\rotatebox{90}{
\begin{tabular}{lc|ccc|ccccc}
\toprule
\textbf{Model} & \textbf{Emb.} & \multicolumn{3}{c|}{\textbf{SOP}, \textbf{CMC metric}} & \multicolumn{5}{c}{\textbf{In-Shop}, \textbf{CMC metric}} \\
& & \textbf{@1} & \textbf{@10} & \textbf{@100} & \textbf{@1} & \textbf{@10} & \textbf{@20} & \textbf{@30} & \textbf{@100} \\\midrule
\multicolumn{10}{c}{\textbf{Embedding-based retrieval}} \\\midrule
HypVit~\cite{9880306} & 128 & 85.5 & 94.9 & \textbf{98.1} & \textbf{92.7} & 98.4 & 98.9 & 99.1 & --- \\
HypVit~\cite{9880306} & 384 & 85.9 & 94.9 & \textbf{98.1} & 92.5 & 98.3 & 98.8 & 99.1 & --- \\
Hyp-DINO~\cite{9880306} & 128 & 84.6 & 94.1 & 97.7 & 92.6 & 98.4 & 99.0 & 99.2 & --- \\
Hyp-DINO~\cite{9880306} & 384 & 85.1 & 94.4 & 97.8 & 92.4 & 98.4 & 98.9 & 99.1 & --- \\
Surrogate recall, ViTs16 backbone~\cite{patel2022recall}  & 384 & 85.6 & 94.8 & 98.0 & --- & --- & --- & --- & --- \\
$\text{IRT}_{\text{R}}$ based on DeiT-s~\cite{DBLP:journals/corr/abs-2102-05644} & 384 & 84.2 & 93.7 & 97.3 & 91.9 & 98.1 & 98.7 & 98.9 & --- \\
ROADMAP (on DeiT-s)~\cite{NEURIPS2021_c622c085} & 384 & 86.0 & 94.4 & 97.6 & --- & --- & --- & --- & --- \\
ViT-Triplet & 384 & \textbf{86.5} & \textbf{95.2} & \textbf{98.1} & 92.1 & \textbf{98.5} & \textbf{99.1} & \textbf{99.3} & 99.7 \\\midrule
\multicolumn{10}{c}{\textbf{Reranking approaches}} \\\midrule
Reranking Transformers (frozen)~\cite{DBLP:conf/iccv/TanYO21} & & 81.8 & 92.4 & 96.6 & --- & --- & --- & --- & --- \\
Reranking Transformers (finetuned)~\cite{DBLP:conf/iccv/TanYO21} & & 84.5 & 93.2 & 96.6 & --- & --- & --- & --- & --- \\
STIR (ViT-Triplet), $n=5$ & 384 & 88.1 & \textbf{95.3} & \textbf{98.1} & 94.9 & \textbf{98.5} & \textbf{99.1} & \textbf{99.3} & 99.7 \\
STIR-Symmetric (ViT-Triplet), $n=5$ & 384 & \textbf{88.3} & \textbf{95.3} & \textbf{98.1} & \textbf{95.0} & \textbf{98.5} & \textbf{99.1} & \textbf{99.3} & 99.7 \\
\bottomrule
\end{tabular}
}
\end{table*}

\begin{table}[t]\centering
\caption{Mean average precision on SOP and In-Shop datasets.}\label{tbl:map}
\setlength{\tabcolsep}{4pt}
\begin{tabular}{lc|cc|cc}
\toprule
\textbf{Model} & \textbf{Emb.} & \multicolumn{2}{c|}{\textbf{SOP}, \textbf{mAP}} & \multicolumn{2}{c}{\textbf{In-Shop}, \textbf{mAP}} \\
& & \textbf{@5} & \textbf{@10} & \textbf{@5} & \textbf{@10} \\\midrule
ViT-Triplet & 384 & 87.6 & 85.1 & 91.6 & 88.4 \\
STIR, $n=5$ & 384 & 89.4 & 86.5 & 94.8 & 91.0 \\
STIR-Symmetric, $n=5$ & 384 & \textbf{89.5} & \textbf{86.6} & \textbf{95.0} & \textbf{91.2} \\
\bottomrule
\end{tabular}
\end{table}

\begin{table}[t]\centering
\caption{Ablation study for STIR, In-Shop dataset.}\label{tbl:stir}
\setlength{\tabcolsep}{4pt}
\begin{tabular}{l|cc|cc}
\toprule
\textbf{Model} & \textbf{CMC@1} & \textbf{CMC@10} & \textbf{mAP@5} & \textbf{mAP@10} \\\midrule
STIR, $n=3$ & 94.4 & \textbf{98.5} & 93.1 & 89.6 \\
STIR, $n=5$ & \textbf{94.9} & \textbf{98.5} & \textbf{94.8} & 91.0 \\
STIR, $n=7$ & \textbf{94.9} & \textbf{98.5} & 94.7 & 92.0 \\
STIR, $n=9$ & \textbf{94.9} & \textbf{98.5} & 94.5 & \textbf{92.7} \\
\bottomrule
\end{tabular}
\end{table}

Table~\ref{tbl:stir} presents the results of our ablation study for STIR variations differing by the number $n$ of the results they rerank. Since STIR requires to run a ViT-based model for each query-gallery pair, it is a relatively heavyweight approach to postprocessing so we limit the comparison to small values of $n$. We see that the CMC@1 (Recall@1) metric saturates quickly as we increase $n$. Table~\ref{tbl:stir} also shows the advantages of mAP in this case: it is a holistic metric that improves as positive samples move closer to the top of the list so mAP@5 and mAP@10 can be used to detect improvements, while CMC@10, naturally, remains unchanged when we rerank top results for $n\le 10$.

\subsection{Qualitative analysis}

\begin{figure}[!t]\centering
\begin{tabular}{c}
    \includegraphics[width=.9\linewidth]{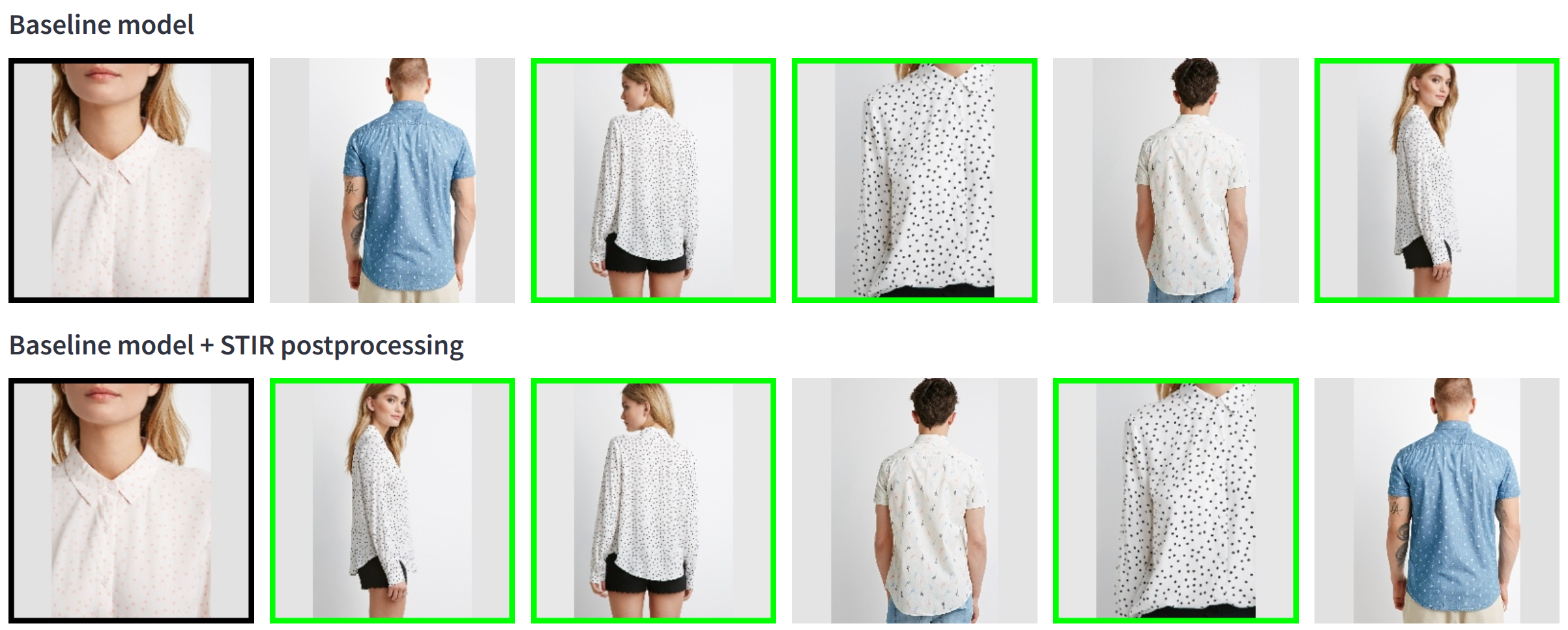}
    \\ (a) \\
    \includegraphics[width=.9\linewidth]{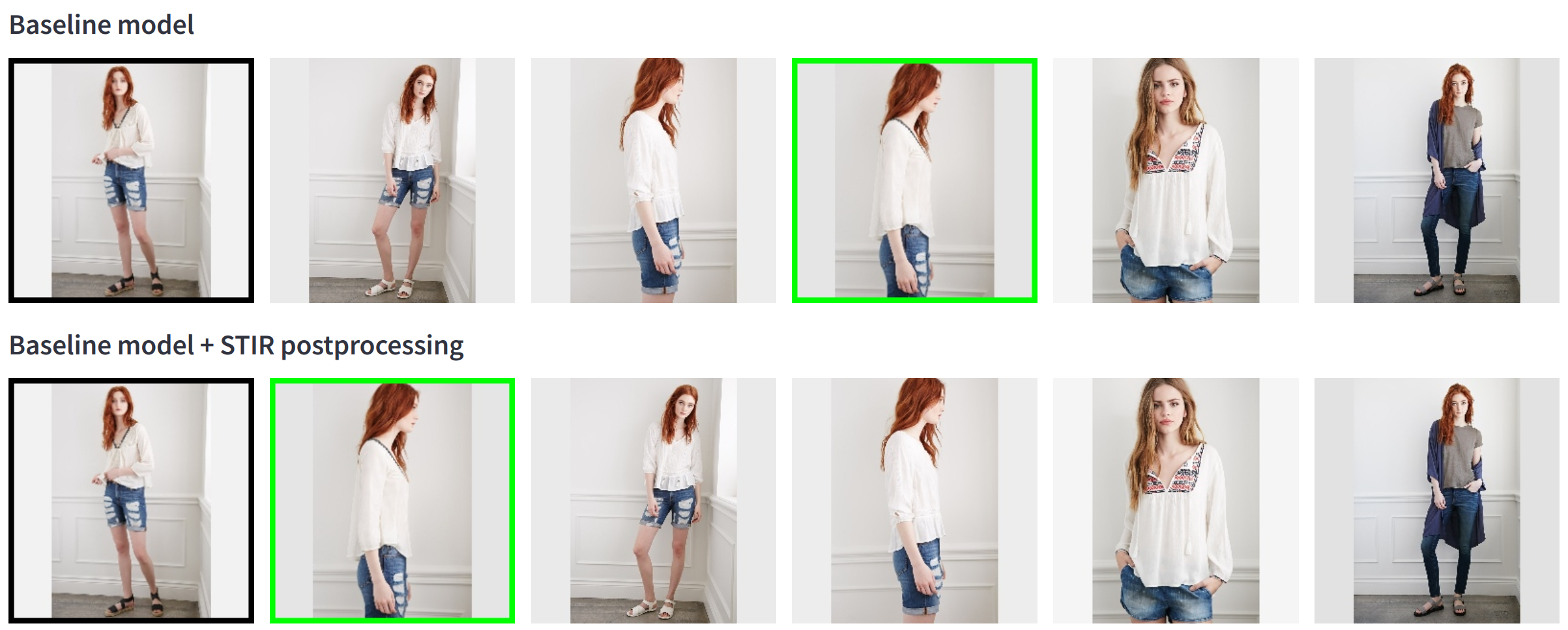}
    \\ (b) \\
    \includegraphics[width=.9\linewidth]{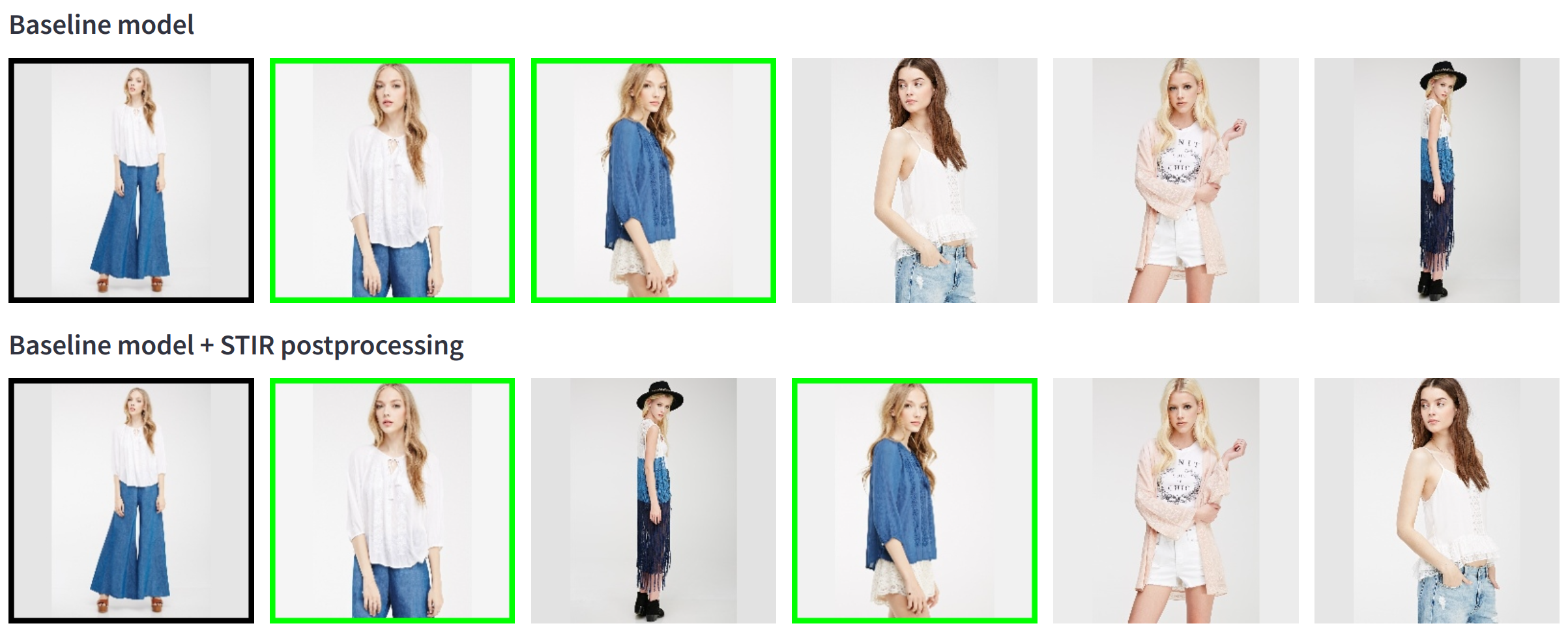}
    \\ (c) 
\end{tabular}
    \caption{STIR postprocessing examples from the interactive demo.}\label{fig:demo}
\end{figure}

Figure~\ref{fig:demo} shows several reranking examples from the InShop dataset as shown in our interactive demo of STIR\footnote{https://dapladoc-oml-postprocessing-demo-srcappmain-pfh2g0.streamlit.app/}. Fig.~\ref{fig:demo}a and Fig.~\ref{fig:demo}b show two results where the reranking improves the results according to the ground truth labeled in the test set; in particular, in both cases the best (top-1) result has been corrected from wrong to right.

Fig.~\ref{fig:demo}c shows a result where STIR reranking actually makes the output worse according to the ground truth labeling. Note, however, that the ground truth results in this case are problematic themselves: they deal only with the shirt of the model while the query clearly shows both the shirt and jeans that do not match in the second ``correct'' answer. Unfortunately, such ambiguous results are encountered in existing datasets quite often, so we note this as a direction for further improvement that might help the entire field of image retrieval. Note also that the InShop dataset is supposed to care about the cut and fashion of a clothing item rather than color, so the model is supposed to retrieve the same item in different colors as well, which often increases the ambiguity.

\section{Conclusion}\label{sec:concl}

In this work, we have presented a simple \emph{ViT-Triplet} model that uses the ViT backbone and the triplet loss and have shown that it consistently reaches or exceeds state of the art results in image retrieval. Thus, a straightforward solution with the best available backbone and a well-tuned training process still remains at the state of the art level in image retrieval.
Moreover, we have presented a postprocessing approach called STIR that reranks top results by an additional pass of ViT over concatenated query and gallery images; STIR is a heavyweight postprocessing method aimed at improving the top of the list. Our experimental study on SOP and In-Shop datasets has shown that STIR can indeed significantly improve retrieval results. We also release a library that implements our methods and can reproduce all our results\footnote{https://github.com/OML-Team/open-metric-learning}.

We note several directions for further work. First, we only consider direct query-to-gallery interactions, while gallery-to-gallery interactions are left indirect. Second, STIR processes the original concatenated images, which makes it relatively slow, and there may be at least two ways to address the problem. First, our current backbone is ViT, which has quadratic complexity with respect to image size; replacing it with an architecture such as the Swin Transformer~\cite{DBLP:journals/corr/abs-2103-14030} that would reduce this complexity to linear may significantly speed up postprocessing. Second, one can replace original images with descriptors obtained from intermediate layers of the feature extractor during the first stage; much of the semantic information is already contained in these features but this would require a different architecture, which we leave for future work.

\newcommand{\etalchar}[1]{$^{#1}$}



\end{document}